# Arithmetic Circuits and the Hadamard Product of Polynomials

V. Arvind, Pushkar S. Joglekar, Srikanth Srinivasan

Institute of Mathematical Sciences
C.I.T Campus,Chennai 600 113, India
{arvind,pushkar,srikanth}@imsc.res.in

**Abstract.** Motivated by the Hadamard product of matrices we define the Hadamard product of multivariate polynomials and study its arithmetic circuit and branching program complexity. We also give applications and connections to polynomial identity testing. Our main results are the following.
- We show that noncommutative polynomial identity testing for algebraic branching programs over rationals is complete for the logspace counting class $C_=L$, and over fields of characteristic $p$ the problem is in $\text{Mod}_p L/\text{poly}$.
- We show an exponential lower bound for expressing the Raz-Yehudayoff polynomial as the Hadamard product of two monotone multilinear polynomials. In contrast the Permanent can be expressed as the Hadamard product of two monotone multilinear formulas of quadratic size.

## 1 Introduction

In this paper we define the *Hadamard product* of two polynomials $f$ and $g$ in $\mathbb{F}\langle X \rangle$ and study its expressive power and applications to the complexity of arithmetic circuits and algebraic branching programs. We also apply it to give a fairly tight characterization of polynomial identity testing for algebraic branching programs over the field of rationals.

Suppose $X = \{x_1, x_2, \cdots, x_n\}$ is a set of $n$ noncommuting variables. The free monoid $X^*$ consists of all words over these variables. For a field $\mathbb{F}$ let $\mathbb{F}\langle x_1, x_2, \cdots, x_n \rangle$ denote the free noncommutative polynomial ring over $\mathbb{F}$ generated by the variables in $X$. Thus, the polynomials in this ring are $\mathbb{F}$-linear combinations of words over $X$. For a given polynomial $f \in \mathbb{F}\langle X \rangle$, let $\text{mon}(f) = \{m \in X^* \mid m \text{ is a nonzero monomial in } f\}$. If $X = \{x_1, x_2, \cdots, x_n\}$ is a set of $n$ *commuting variables* then $\mathbb{F}[X]$ denotes the commutative polynomial ring with coefficients from $\mathbb{F}$.

Motivated by the well-known Hadamard product of matrices (see e.g. [Bh97]) we define the Hadamard product of polynomials.

**Definition 1.** *Let $f, g \in \mathbb{F}\langle X \rangle$ where $X = \{x_1, x_2, \cdots, x_n\}$. The* Hadamard product *of $f$ and $g$, denoted $f \circ g$, is the polynomial $f \circ g = \sum_m a_m b_m m$, where $f = \sum_m a_m m$ and $g = \sum_m b_m m$, where the sums index over monomials $m$.*

**Complexity theory preliminaries** We recall some definitions of logspace counting classes from [AO96]. Let Ł denote the class of languages accepted by deterministic logspace machines.

GapL is the class of functions $f : \Sigma^* \to \mathbb{Z}$, for which there is a logspace bounded NDTM $M$ such that for each input $x \in \Sigma^*$, we have $f(x) = acc_M(x) - rej_M(x)$, where $acc_M(x)$ and $rej_M(x)$ are the number of accepting and rejecting paths of $M$ on input $x$, respectively.

A language $L$ is in $C_=L$ if there exists a function $f \in$ GapL such that $x \in L$ if and only if $f(x) = 0$. For a prime $p$, a language $L$ is in the complexity class $Mod_pL$ if there exists a function $f \in$ GapL such that $x \in L$ if and only if $f(x) = 0 (\mod p)$.

It is shown in [AO96] that checking if an integer matrix is singular is complete for $C_=L$ with respect to logspace many-one reductions. The same problem is known to be complete for $Mod_pL$ over a field of characteristic $p$. It is useful to recall that both $C_=L$ and $Mod_pL$ are contained in $TC^1$ (which, in turn, is contained in $NC^2$).

An *Algebraic Branching Program* (ABP) [N91,RS05] over a field $\mathbb{F}$ and variables $x_1, x_2, \cdots, x_n$ is a *layered* directed acyclic graph with one *source* vertex of indegree zero and one *sink* vertex of outdegree zero. Let the layers be numbered $0, 1, \cdots, d$. The source and sink are the unique layer $0$ and layer $d$ vertices, respectively. Edges only go from layer $i$ to $i+1$ for each $i$. Each edge in the ABP is labeled with a linear form over $\mathbb{F}$ in the input variables. Each source to sink path in the ABP computes the product of the linear forms labelling the edges on the path, and the sum of these polynomials over all source to sink paths is the polynomial computed by the ABP. The size of the ABP is the number of vertices.

**Main results.** We show that the *noncommutative* branching program complexity of the Hadamard product $f \circ g$ is upper bounded by the product of the branching program sizes for $f$ and $g$. This upper bound is natural because we know from Nisan's seminal work [N91] that the algebraic branching program (ABP) complexity $B(f)$ is well characterized by the ranks of its "communication" matrices $M_k(f)$, and the rank of Hadamard product $A \circ B$ of two matrices $A$ and $B$ is upper bounded by the product of their ranks. Our proof is constructive: we give a deterministic logspace algorithm for computing an ABP for $f \circ g$.

We then apply this result to polynomial identity testing. It is shown by Raz and Shpilka [RS05] that polynomial identity testing of noncommutative ABPs can be done in deterministic polynomial time. A simple divide and conquer algorithm can be easily designed to show that the problem is in deterministic $NC^3$. What then is the precise complexity of polynomial identity testing for noncommutative ABPs? For noncommutative ABPs over *rationals* we give a tight characterization by showing that the problem is $C_=L$-complete. We prove this result using the result on Hadamard product of ABPs explained above.

For noncommutative ABPs over a finite field of characteristic $p$, we show that identity testing is in the nonuniform class $\text{Mod}_p\text{L}/\text{poly}$ (more precisely, in randomized $\text{Mod}_p\text{L}$). Furthermore, the problem turns out to be hard (w.r.t. logspace many-one reductions) for both NL and $\text{Mod}_p\text{L}$. Hence, it is not likely to be easy to improve this upper bound unconditionally to $\text{Mod}_p\text{L}$ (it would imply that NL is contained in $\text{Mod}_p\text{L}$). However, under a hardness assumption we can apply standard arguments [ARZ99,KvM02] to derandomize this algorithm and put the problem in $\text{Mod}_p\text{L}$.

In Section 4 we consider the Hadamard product for commutative polynomials. We show an exponential lower bound for expressing the Raz-Yehudayoff polynomial [RY08] as the Hadamard product of two monotone multilinear polynomials. In contrast the Permanent can be expressed as the Hadamard product of two monotone multilinear formulas of quadratic size.

## 2 The Hadamard Product

Let $f, g \in \mathbb{F}\langle X \rangle$ where $X = \{x_1, x_2, \cdots, x_n\}$. Clearly, $\text{mon}(f \circ g) = \text{mon}(f) \cap \text{mon}(g)$. Thus, the Hadamard product can be seen as an algebraic version of the intersection of formal languages. Our definition of the Hadamard product of polynomials is actually motivated by the well-known Hadamard product $A \circ B$ of two $m \times n$ matrices $A$ and $B$. We recall the following well-known bound for the rank of the Hadamard product.

**Proposition 1.** *Let $A$ and $B$ be $m \times n$ matrices over a field $\mathbb{F}$. Then $\text{rank}(A \circ B) \leq \text{rank}(A)\text{rank}(B)$.*

It is known from Nisan's work [N91] that the ABP complexity $B(f)$ of a polynomial $f \in \mathbb{F}\langle X \rangle$ is closely connected with the ranks of the communication matrices $M_k(f)$, where $M_k(f)$ has its rows indexed by degree $k$ monomials and columns by degree $d - k$ monomials and the $(m, m')^{th}$ entry of $M_k(f)$ is the coefficient of $mm'$ in $f$. Nisan showed that $B(f) = \sum_k \text{rank}(M_k(f))$. Indeed, Nisan's result and the above proposition easily imply the following bound on the ABP complexity of $f \circ g$.

**Lemma 1.** *For $f, g \in \mathbb{F}\langle X \rangle$ we have $B(f \circ g) \leq B(f)B(g)$.*

*Proof.* By Nisan's result $B(f \circ g) = \sum_k \text{rank}(M_k(f \circ g))$. The above proposition implies

$$\sum_k \text{rank}(M_k(f \circ g)) \leq \sum_k \text{rank}(M_k(f))\text{rank}(M_k(g)) \leq (\sum_k \text{rank}(M_k(f)))(\sum_k \text{rank}(M_k(g))),$$

and the claim follows. ∎

We now show an algorithmic version of this upper bound.

**Theorem 1.** *Let $P$ and $Q$ be two given ABP's computing polynomials $f$ and $g$ in $\mathbb{F}\langle x_1, x_2, \ldots, x_n\rangle$, respectively. Then there is a deterministic polynomial-time algorithm that will output an ABP $R$ for the polynomial $f \circ g$ such that the size of $R$ is a constant multiple of the product of the sizes of $P$ and $Q$. (Indeed, $R$ can be computed in deterministic logspace.)*

*Proof.* Let $f_i$ and $g_i$ denote the $i^{th}$ homogeneous parts of $f$ and $g$ respectively. Then $f = \sum_{i=0}^{d} f_i$ and $g = \sum_{i=0}^{d} g_i$. Since the Hadamard product is distributive over addition and $f_i \circ g_j = 0$ for $i \neq j$ we have $f \circ g = \sum_{i=0}^{d} f_i \circ g_i$. Thus, we can assume that both $P$ and $Q$ are homogeneous ABP's of degree $d$. Otherwise, we can easily construct an ABP to compute $f_i \circ g_i$ separately for each $i$ and put them together. Note that we can easily compute ABPs for $f_i$ and $g_i$ in logspace given as input the ABPs for $f$ and $g$.

By allowing parallel edges between nodes of $P$ and $Q$ we can assume that the labels associated with each edge in an ABP is either $0$ or $\alpha x_i$ for some variable $x_i$ and scalar $\alpha \in \mathbb{F}$. Let $s_1$ and $s_2$ bound the number of nodes in each layer of $P$ and $Q$ respectively. Denote the $j^{th}$ node in layer $i$ by $\langle i, j \rangle$ for ABPs $P$ and $Q$. Now we describe the construction of the ABP $R$ for computing the polynomial $f \circ g$. Each layer $i$, $1 \leq i \leq d$ of $R$ will have $s_1 \cdot s_2$ nodes, with node labeled $\langle i, a, b \rangle$ corresponding to the node $\langle i, a \rangle$ of $P$ and the node $\langle i, b \rangle$ of $Q$. We can assume there is an edge from every node in layer $i$ to every node in layer $i+1$ for both ABPs. For, if there is no such edge we can always include it with label $0$.

In the new ABP $R$ we put an edge from $\langle i, a, b \rangle$ to $\langle i+1, c, e \rangle$ with label $\alpha\beta x_t$ if and only if there is an edge from node $\langle i, a \rangle$ to $\langle i+1, c \rangle$ with label $\alpha x_t$ in $P$ and an edge from $\langle i, b \rangle$ to $\langle i+1, e \rangle$ with label $\beta x_t$ in ABP $Q$. Let $\langle 0, a, b \rangle$ and $\langle d, c, e \rangle$ denote the source and the sink nodes of ABP $R$, where $\langle 0, a \rangle, \langle 0, b \rangle$ are the source nodes of $P$ and $Q$, and $\langle d, c \rangle, \langle d, e \rangle$ are the sink nodes of $P$ and $Q$ respectively. It is easy to see that ABP $R$ can be computed in deterministic logspace. Let $h_{\langle i, a, b \rangle}$ denote the polynomial computed at node $\langle i, a, b \rangle$ of ABP $R$. Similarly, let $f_{\langle i, a \rangle}$ and $g_{\langle i, b \rangle}$ denote the polynomials computed at node $\langle i, a \rangle$ of $P$ and node $\langle i, b \rangle$ of $Q$. We can easily check that $h_{\langle i, a, b \rangle} = f_{\langle i, a \rangle} \circ g_{\langle i, b \rangle}$ by an induction argument on the number of layers in the ABPs. It follows from this inductive argument that the ABP $R$ computes the polynomial $f \circ g$ at its sink node. The bound on the size of $R$ also follows easily. ∎

Applying the above theorem we can give a *tight* complexity theoretic upper bound for identity testing of noncommutative ABPs over rationals.

**Theorem 2.** *The problem of polynomial identity testing for noncommutative algebraic branching programs over $\mathbb{Q}$ is in $\text{NC}^2$. More precisely, it complete for the logspace counting class $\text{C}_=\text{L}$ under logspace reductions.*

*Proof.* Let $P$ be the given ABP computing $f \in \mathbb{Q}\langle X \rangle$. We apply the construction of Theorem 1 to compute a polynomial sized ABP $R$ for the Hadamard product $f \circ f$ (i.e. of $f$ with itself). Notice that $f \circ f$ is nonzero iff $f$ is nonzero. Now, we crucially use the fact that $f \circ f$ is a polynomial whose nonzero coefficients are all *positive*. Hence, $f \circ f$ is nonzero iff it evaluates to nonzero on the all 1's input. The problem thus boils down to checking if $R$ evaluates to nonzero on the all 1's input.

By Theorem 1, the ABP $R$ for polynomial $f \circ f$ is computable in deterministic logspace, given as input an ABP for $f$. Furthermore, evaluating the ABP $R$ on the all 1's input can be easily converted to iterated integer matrix multiplication (one matrix for each layer of the ABP), and checking if $R$ evaluates to nonzero can be done by checking if a specific entry of the product matrix is nonzero. It is well known that checking if a specific entry of an iterated integer matrix product is zero is in the logspace counting class $C_=L$ (e.g. see [AO96,ABO99]). However, $C_=L$ is contained in $NC^2$, in fact in $TC^1$.

We now argue the hardness of this problem for $C_=L$. The problem of checking if an integer matrix $A$ is singular is well known to be complete for $C_=L$ under deterministic logspace reductions. The standard GapL algorithm for computing $\det(A)$ [T91] can be converted to an ABP $P_A$ which will compute $\det(A)$.[1] Hence the ABP $P_A$ computes the identically zero polynomial iff $A$ is singular. Putting it all together, it follows that identity testing of noncommutative ABPs over rationals is complete for the class $C_=L$. ∎

**An iterative matrix product problem** Suppose $B$ is a noncommutative ABP computing a homogeneous polynomial in $\mathbb{F}\langle X \rangle$ of degree $d$, where each edge of the ABP is labeled by a homogeneous linear form in variables from $X$.

Let $n_\ell$ denote the number of nodes of $B$ in layer $\ell$, $0 \leq \ell \leq d$. For each $x_i$ and layer $\ell$, we associate an $n_\ell \times n_{\ell+1}$ matrix $A_{i,\ell}$ where the $(k,j)^{th}$ entry of matrix $A_{i,\ell}$ is the coefficient of $x_i$ in the linear form associated with the $(v_k, u_j)$ edge in the ABP $B$. Here $v_k$ is the $k^{th}$ node in layer $\ell$ and $u_j$ the $j^{th}$ node in the layer $\ell + 1$. The following claim is easy to see and relates these matrices to the ABP $B$.

*Claim.* The coefficient of any degree $d$ monomial $x_{i_1} x_{i_2} \cdots x_{i_d}$ in the polynomial computed by the ABP $B$ is the matrix product $A_{i_1,0} A_{i_2,1} \cdots A_{i_d,d-1}$ (which is a scalar since $A_{i_1,0}$ is a row and $A_{i_d,d-1}$ is a column).

Let $i$ and $j$ be any two nodes in the ABP $B$. We denote by $B(i,j)$ the algebraic branching program obtained from the ABP $B$ by designating node $i$ in $B$ as the source node and node $j$ as the sink node. Clearly, $B(i,j)$ computes a homogeneous polynomial of degree $b - a$ if $i$ appears in layer $a$ and $j$ in layer $b$.

---

[1] Notice that the polynomial computed by the ABP $P_A$ is a constant since $P_A$ has only constants and no variables.

For layers $a, b$, $0 \leq a < b \leq d$ let $t = b - a$ and $P(a,b) = \{A_{s_1,a} A_{s_2,a+1} \ldots A_{s_t,b-1} | 1 \leq s_j \leq n,\text{ for } 1 \leq j \leq t\}$. $P(a,b)$ consists of $n_a \times n_b$ matrices. Thus the dimension of the linear space spanned by $P(a,b)$ is bounded by $n_a n_b$. It follows from Claim 2 that the linear span of $P(a,b)$ is the zero space iff the polynomial computed by ABP $B(i,j)$ is identically zero for every $1 \leq i \leq n_a$ and $1 \leq j \leq n_b$.

Thus, it suffices to compute a basis for the space spanned by matrices in $P(0,d)$ to check whether the polynomial computed by $B$ is identically zero. We can easily give a deterministic $\text{NC}^3$ algorithm for this problem over any field $\mathbb{F}$: First recursively compute bases $M_1$ and $M_2$ for the space spanned by matrices in $P(0, d/2)$ and $P(d/2+1, d)$ respectively. From bases $M_1$ and $M_2$ we can compute in deterministic $\text{NC}^2$ a basis $M$ for space spanned by matrices in $P(0,d)$ as follows. We compute the set $S$ of pairwise products of matrices in $M_1$ and $M_2$ and then we can compute a maximal linearly independent subset of $S$ in $\text{NC}^2$ (see e.g. [ABO99]). This gives an easy $\text{NC}^3$ algorithm to compute a basis for the linear span of $P(0,d)$. This proves the following.

**Proposition 2.** *The problem of polynomial identity testing for noncommutative algebraic branching programs over any field (in particular, finite fields $\mathbb{F}$) is in deterministic $\text{NC}^3$.*

Can we give a tight complexity characterization for identity testing of noncommutative ABPs over finite fields? We show that the problem is in nonuniform $\text{Mod}_p\text{L}$ and is hard for $\text{Mod}_p\text{L}$ under logspace reductions. Furthermore, the problem is hard for NL. Hence, it appears difficult to improve the upper bound to uniform $\text{Mod}_p\text{L}$ (as NL is not known to be contained in uniform $\text{Mod}_p\text{L}$).

**Theorem 3.** *The problem of polynomial identity testing for noncommutative algebraic branching programs over a finite field $\mathbb{F}$ of characteristic $p$ is in $\text{Mod}_p\text{L}/\text{poly}$.*

*Proof.* Consider a new ABP $B'$ in which we replace the variables $x_i$, $1 \leq i \leq n$ appearing in the linear form associated with an edge from some node in layer $l$ to a node in layer $l+1$ of ABP $B$ by new variable $x_{i,l}$, for layers $l = 0, 1, \ldots, d-1$. Let $g \in \mathbb{F}[X]$ denotes the polynomial computed by ABP $B'$ in *commuting* variables $x_{i,l}, 1 \leq i \leq n, 1 \leq l < d$. It is easy to see that the commutative polynomial $g \in \mathbb{F}[X]$ is identically zero iff the noncommutative polynomial $f \in \mathbb{F}\langle X \rangle$ computed by ABP $B$ is identically zero. Now, we can apply the standard Schwartz-Zippel lemma to check if $g$ is identically zero by substituting random values for the variables $x_{i,l}$ from $\mathbb{F}$ (or a suitable finite extension of $\mathbb{F}$). After substitution of field elements, we are left with an iterated matrix product over a field of characteristic $p$ which can be done in $\text{Mod}_p\text{L}$. This gives us a randomized $\text{Mod}_p\text{L}$ algorithm. By standard amplification it follows that the problem is in $\text{Mod}_p\text{L}/\text{poly}$. ∎

Next we show that identity testing noncommutative ABPs over any field is hard for NL by a reduction from directed graph reachability. Let $(G, s, t)$ be a reachability instance. Without loss of generality, we assume that $G$ is a layered directed acyclic graph. The graph $G$ defines an ABP with source $s$ and sink $t$ as follows: label each edge $e$ in $G$ with a *distinct* variable $x_e$ and for each absent edge put the label 0. The polynomial computed by the ABP is nonzero if and only if there is a directed $s$-$t$ path in $G$.

**Theorem 4.** *The problem of polynomial identity testing for noncommutative algebraic branching programs over any field is hard for* NL.

## 3  Hadamard product of noncommutative circuits

Analogous to Theorem 1 we show that $f \circ g$ has small circuits if $f$ has a small circuit and $g$ has a small ABP.

**Theorem 5.** *Let $f, h \in \mathbb{F}\langle x_1, x_2, \cdots, x_n \rangle$ be given by a degree $d$ circuit $C$ and a degree $d$ ABP $P$ respectively, where $d = O(n^{O(1)})$. Then we can compute in polynomial time a circuit $C'$ that computes $f \circ h$ where the size of $C'$ is polynomially bounded in the sizes of $C$ and $P$.*

*Proof.* As in the proof of Theorem 1 we can assume that both $f$ and $h$ are homogeneous polynomials of degree $d$. Let $f_g$ denote the polynomial computed at gate $g$ of circuit $C$. Let $w$ bound the number of nodes in any layer of $P$. Let $\langle i, a \rangle$ denote the $a^{th}$ node in the $i^{th}$ layer of $P$ for $0 \leq i \leq d, 1 \leq a \leq w$. Let $h_{(i,a),(j,b)}$ denote the polynomial computed by ABP $P'$, where $P'$ is same as $P$ but with source node $\langle i, a \rangle$ and sink node $\langle j, b \rangle$. We now describe the circuit $C'$ computing the polynomial $f \circ h$. In $C'$ we have gates $\langle g, l, (i, a), (i+l, b) \rangle$ for $0 \leq l \leq d, 0 \leq i \leq d, 1 \leq a, b \leq w$ associated with each gate $g$ of $C$, such that at the gate $\langle g, l, (i, a), (i+l, b) \rangle$ the circuit $C'$ computes

$$r^{\langle g,l \rangle}_{(i,a),(i+l,b)} = f_{\langle g,l \rangle} \circ h_{(i,a),(i+l,b)} \tag{1}$$

where $f_{\langle g,l \rangle}$ denotes the degree $l$ homogeneous component of the polynomial $f_g$.

If $g$ is a + gate of $C$ with input gates $g_1, g_2$ so that $f_g = f_{g_1} + f_{g_2}$, we have $r^{\langle g,l \rangle}_{(i,a),(i+l,b)} = r^{\langle g_1,l \rangle}_{(i,a),(i+l,b)} + r^{\langle g_2,l \rangle}_{(i,a),(i+l,b)}$, for $0 \leq l \leq d, 0 \leq i \leq d, 1 \leq a, b \leq w$. In other words, $\langle g, l, (i, a), (i+l, b) \rangle$ is a + gate in $C'$ with input gates $\langle g_1, l, (i, a), (i+l, b) \rangle$ and $\langle g_2, l, (i, a), (i+l, b) \rangle$. If $g$ is a $\times$ gate in $C$ we will have

$$r^{\langle g,l \rangle}_{(i,a),(i+l,b)} = \sum_{j=0}^{l} \sum_{t=1}^{w} r^{\langle g_1,j \rangle}_{(i,a),(i+j,t)} \cdot r^{\langle g_2,l-j \rangle}_{(i+j,t),(i+l,b)} \tag{2}$$

The above formula is easily computable by a small subcircuit. The output gate of $C'$ will be $\langle g, d, (0, 1), (d, 1) \rangle$, where $g$ is the output gate of $C$, and $(0, 1)$ and $(d, 1)$ are the source and the sink of the ABP $P$ respectively. This is the description of the circuit $C'$. We inductively argue that gate $\langle g, l, (i, a), (i+l, b)\rangle$ of $C'$ computes the polynomial $f_{\langle g,l \rangle} \circ h_{(i,a),(i+l,b)}$. If $g$ is a $+$ gate of $C$ the claim is obvious. Suppose $g$ is a $\times$ gate of $C$ with inputs $g_1, g_2$ such that $f_g = f_{g_1} \cdot f_{g_2}$. Inductively assume that the claim holds for the gates $g_1$ and $g_2$. Then we have $f_{\langle g,l \rangle} = \sum_{i=0}^{l} f_{\langle g_1,i \rangle} \cdot f_{\langle g_2,l-i \rangle}$. Hence, it easily follows that

$$f_{\langle g,l \rangle} \circ h_{(i,a),(i+l,b)} = \sum_{j=0}^{l} (f_{\langle g_1,j \rangle} \cdot f_{\langle g_2,l-j \rangle} \circ h_{(i,a),(i+l,b)})$$

$$= \sum_{j=0}^{l} \sum_{t=1}^{w} f_{\langle g_1,j \rangle} \cdot f_{\langle g_2,l-j \rangle} \circ h_{(i,a),(i+j,t)} \cdot h_{(i+j,t),(i+l,b)}$$

$$= \sum_{j=0}^{l} \sum_{t=1}^{w} (f_{\langle g_1,j \rangle} \circ h_{(i,a),(i+j,t)}) \cdot (f_{\langle g_2,l-j \rangle} \circ h_{(i+j,t),(i+l,b)})$$

By induction hypothesis we have $r^{\langle g_1,j \rangle}_{(i,a),(i+j,t)} = f_{\langle g_1,j \rangle} \circ h_{(i,a),(i+j,t)}$ and $r^{g_2,l-j}_{(i+j,t),(i+l,b)} = f_{\langle g_2,l-j \rangle} \circ h_{(i+j,t),(i+l,b)}$. Now, from Equation 2 it is easy to obtain the desired Equation 1. Therefore, at the output gate $\langle g, d, (0,1), (d,1)\rangle$ the circuit $C'$ computes $f \circ h$. The size of $C'$ is bounded by a polynomial in the sizes of $C$ and $P$. ∎

On the other hand, suppose $f$ and $g$ individually have small circuit complexity. Does $f \circ g$ have small circuit complexity? Can we compute such a circuit for $f \circ g$ from circuits for $f$ and $g$? We first consider these questions for monotone circuits. It is useful to understand the connection between monotone noncommutative circuits and context-free grammars. We recall the following definition.

**Definition 2.** *We call a context-free grammar $G = (V, T, P, S)$ an* acyclic CFG *if for any nonterminal $A \in V$ there does not exist any derivation of the form $A \Rightarrow^* uAw$, and for each production $A \Rightarrow \beta$ we have $|\beta| \leq 2$.*

The size $size(G)$ of an acyclic CFG $G = (V, T, P, S)$ is defined as $|V| + |T| + \text{size}(P)$, where $V$, $T$, and $P$ are the sets of variables, terminals, and production rules. We note the following easy proposition that relates acyclic CFGs to monotone noncommutative circuits over $X$. Proof of the Proposition 3 is in the Appendix.

**Proposition 3.** *Let $C$ be a monotone circuit of size $s$ computing a polynomial $f \in \mathbb{Q}\langle X \rangle$. Then there is an acyclic CFG $G$ for $\text{mon}(f)$ with $\text{size}(G) = O(s)$. Conversely, if $G$ is an acyclic CFG of size $s$ computing some finite set $L \subset X^*$ of monomials over $X$, there exists a monotone circuit of size $O(s)$ that computes a polynomial $\sum_{m \in L} a_m m \in \mathbb{Q}\langle X \rangle$, where the positive integer $a_m$ is the number of derivation trees for $m$ in the grammar $G$.*

**Theorem 6.** *There are monotone circuits $C$ and $C'$ computing polynomials $f$ and $g$ in $\mathbb{Q}\langle X \rangle$ respectively, such that the polynomial $f \circ g$ requires monotone circuits of size exponential in $|X|$, $\text{size}(C)$, and $\text{size}(C')$.*

*Proof.* Let $X = \{x_1, \cdots, x_n\}$. Define the finite language $L_1 = \{zww^r \mid z, w \in X^*, |z| = |w| = n\}$ and the corresponding polynomial $f = \sum_{m_\alpha \in L_1} m_\alpha$. Similarly let $L_2 = \{ww^r z \mid z, w \in X^*, |z| = |w| = n\}$, and the corresponding polynomial $g = \sum_{m_\alpha \in L_2} m_\alpha$. It is easy to see that there are poly$(n)$ size *unambiguous* acyclic CFGs for $L_1$ and $L_2$. Hence, by Proposition 3 there are monotone circuits $C_1$ and $C_2$ of size poly$(n)$ such that $C_1$ computes polynomial $f$ and $C_2$ computes polynomial $g$. We first show that the finite language $L_1 \cap L_2$ cannot be generated by any acyclic CFG of size $2^{o(n \lg n)}$. Assume to the contrary that there is an acyclic CFG $G = (V, T, P, S)$ for $L_1 \cap L_2$ of size $2^{o(n \lg n)}$. Notice that $L_1 \cap L_2 = \{t \mid t = ww^r w, w \in X^*, |w| = n\}$.

Consider any derivation tree $T'$ for a word $ww^r w = w_1 w_2 \ldots w_n w_n w_{n-1} \ldots w_2 w_1 w_1 \ldots w_n$. Starting from the root of the binary tree $T'$, we traverse down the tree always picking the child with larger yield. Clearly, there must be a nonterminal $A \in V$ in this path of the derivation tree such that $A \Rightarrow^* u$, $u \in X^*$ and $n \leq |u| < 2n$. Crucially, note that any word that $A$ generates must have same length since every word generated by the grammar $G$ is in $L_1 \cap L_2$ and hence of length $3n$. Let $ww^r w = s_1 u s_2$ where $|s_1| = k$. As $|u| < 2n$, the string $s_1 s_2$ completely determines the string $ww^r w$. Hence, the nonterminal $A$ can derive at most one string $u$. Furthermore, this string $u$ can occur in at most $2n$ positions in a string of length $3n$. Notice that for each position in which $u$ can occur it completely determines a string of the form $ww^r w$. Therefore, $A$ can participate in the derivation of at most $2n$ strings from $L_1 \cap L_2$. Since there are $n^n$ distinct words in $L_1 \cap L_2$, it follows that there must be at least $\frac{n^n}{2n}$ *distinct* nonterminals in $V$. This contradicts the size assumption of $G$.

Since $L_1 \cap L_2$ cannot be generated by any acyclic CFG of size $2^{o(n \log n)}$, it follows from Lemma 3 that the polynomial $f \circ g$ can not be computed by any monotone circuit of $2^{o(n \log n)}$ size. ∎

Theorem 6 shows that the Hadamard product of monotone circuits is more expressive than monotone circuits. It raises the question whether the permanent polynomial can be expressed as the Hadamard product of polynomial-size (or even subexponential size) monotone circuits. We note here that the permanent *can* be easily expressed as the Hadamard product of $O(n^3)$ many monotone circuits (in fact, monotone ABPs).

**Theorem 7.** *Suppose there is a deterministic subexponential-time algorithm that takes two circuits as input, computing polynomials $f$ and $g$ in $\mathbb{Q}\langle x_1, \cdots, x_n \rangle$, and outputs a circuit for $f \circ g$. Then either NEXP is not in P/poly or the Permanent does not have polynomial size noncommutative circuits.*

*Proof.* Let $C_1$ be a circuit computing some polynomial $h \in \mathbb{Q}\langle x_1, \ldots, x_n \rangle$. By assumption, we can compute a circuit $C_2$ for $h \circ h$ in subexponential time. Therefore, $h$ is identically zero iff $h \circ h$ is identically zero iff $C_2$ evaluates to 0 on the all 1's input. We can easily check if $C_2$ evaluates to 0 on all 1's input by substitution and evaluation. This gives a deterministic subexponential time algorithm for testing if $h$ is identically zero. By the noncommutative analogue of [KI03], shown in [AMS08], it follows that either NEXP $\not\subset$ P/poly or the Permanent does not have polynomial size noncommutative circuits. ∎

Next, We show that the identity testing problem: given $f, g \in \mathbb{F}\langle X \rangle$ by circuits test if $f \circ g$ is identically zero is coNP hard. The proof of Theorem 8 is given in the Appendix.

**Theorem 8.** *Given two monotone polynomial-degree circuits $C$ and $C'$ computing polynomial $f, g \in \mathbb{Q}\langle X \rangle$ it is coNP-complete to check if $f \circ g$ is identically zero.*

## 4 Hadamard product of monotone multilinear circuits

In this section we study the Hadamard product of *commutative* polynomials (defined as in the noncommutative case). First we introduce some notation useful for this section. Given a polynomial $f \in \mathbb{F}[X]$, and a monomial $m$ over the variables $X$, we define $f(m)$ to be the coefficient of the monomial $m$ in the polynomial $f$.[2] Recall the Definition 1 of the Hadamard product of two polynomials in $\mathbb{F}\langle X \rangle$. We define the Hadamard product in the commutative case analogously. Thus, for polynomials $f, g \in \mathbb{F}[X]$ we have $F(m) = f(m)g(m)$ for any monomial $m$, where $F = f \circ g$.

In this section our interest is the expressive power of the Hadamard product. Can we express a hard explicit polynomial (like the Permanent) as the Hadamard product $f \circ g$ where $f$ and $g$ have small arithmetic circuits? It turns out that we easily can.

**Proposition 4.** *There are multilinear polynomials $f, g \in \mathbb{F}[x_{11}, x_{12}, \cdots, x_{nn}]$ such that both $f$ and $g$ have arithmetic formulas of size $O(n^2)$ and $f \circ g$ is the Permanent polynomial. Furthermore, for $\mathbb{F} = \mathbb{Q}$ these formulas for $f$ and $g$ are monotone.*

*Proof.* Define the polynomials $f$ and $g$ on the variables $\{x_{ij} \mid 1 \leq i, j \leq n\}$ as follows $f = \prod_{i=1}^{n}(\sum_{j=1}^{n} x_{ij})$ and $g = \prod_{j=1}^{n}(\sum_{i=1}^{n} x_{ij})$. Clearly, their Hadamard product is $Perm(x_{11}, \cdots, x_{nn})$. The formulas for $f$ and $g$ over rationals are monotone. ∎

Nevertheless, we will define an explicit monotone multilinear polynomial that cannot be written as the Hadamard product of multilinear polynomials computed by subexponential sized monotone arithmetic circuits. Our construction adapts a result of Raz and

---

[2] There should be no confusion with evaluating the multivariate polynomial $f$ at a point $(a_1, \cdots, a_n)$ as we denote that by $f(a_1, a_2, \cdots, a_n)$.

Yehudayoff [RY08] describing an explicit monotone polynomial that has no monotone arithmetic circuits of size $2^{\epsilon n}$, for some constant $\epsilon > 0$. Our proof closely follows the arguments in [RY08]. Due to lack of space, we provide only proof sketches for several technical statements.

**Definition 3.** *For $\epsilon > 0$, a multilinear polynomial $f \in \mathbb{C}[x_1, \ldots, x_n]$ is an $\epsilon$-product polynomial if there are disjoint sets $A, B \subseteq X = \{x_1, \ldots, x_n\}$ such that $|A| \geq \epsilon n$ and $|B| \geq \epsilon n$ and $f = gh$ where $g \in \mathbb{C}[A]$ and $h \in \mathbb{C}[B]$.*

In the sequel, we often identify a multilinear polynomial $f$ in $\mathbb{C}[X]$ with its coefficients vector (indexed by monomials in the natural lexicographic order). The complex inner product of vectors $w, w' \in \mathbb{C}^k$ is $\langle w, w' \rangle = \sum_i w_i \overline{w'_i}$. Let $\mathcal{M}(X)$ denote the set of multilinear monomials over the variables in $X$.

**Definition 4.** *The correlation of multilinear polynomials $f$ and $g$ in $\mathbb{C}[X]$ is defined as $\mathrm{Corr}(f, g) = |\sum_{m \in \mathcal{M}(X)} f(m)\overline{g(m)}|$. Notice that $\mathrm{Corr}(f, f)$ is the $\ell_2$-norm $\|f\|$ of $f$.*

**The explicit polynomial from [RY08]** The explicit polynomial $F$ we define is essentially the same as the one in [RY08] (the difference is in the constants). Let $s \in \mathbb{N}$ be a constant, to be chosen later and $t = 40s$. Let $n = tp = 40sp$, for a prime $p$, and $X = \{x_1, \ldots, x_n\}$. Partition $X$ into $t$ many sets of variables $X(1), \ldots, X(t)$ with $p$ variables each, where $X(i) = \{x_{(i-1)p+j} \mid j \in [p]\}$.

In poly$(n)$ time we can construct the field $\mathbb{F} = \mathbb{F}_{2^p}$ which is in bijective correspondence with $\{0,1\}^p$. We can assume that $0 \in \mathbb{F}$ is associated with the all 0s vector $0^p$. Fix a nontrivial additive character $\psi$ of $\mathbb{F}$. Since char$(\mathbb{F}) = 2$ we have $\psi(x) = \pm 1$ for all $x \in \mathbb{F}$. Each monomial $m \in \mathcal{M}(X)$ defines a subset $A_m$ of $X$ and is thus represented by its characteristic vector $w \in \{0,1\}^n$. Split $w$ into $t$ blocks $w_1, \ldots, w_t$ of size $p$ each ($w_i$ is the characteristic vector of $A_m \cap X(i)$), and consider the $p$ field elements $y_1(m), y_2(m), \ldots, y_t(m) \in \mathbb{F}$ associated with these strings. The bijection between $\mathbb{F}$ and $\{0,1\}^p$ implies for any $m \in \mathcal{M}(X)$ that $y_i(m) = 0$ iff no variable $x \in X(i)$ appears in $m$.

Let us now define the polynomial $F$. Given a monomial $m \in \mathcal{M}(X)$, we define $F(m)$ to be $\psi(\prod_{i=1}^{t} y_i(m))$. We define a polynomial $f \in \mathbb{F}[X]$ to be *explicit* if the coefficient $f(m)$ of any monomial $m$ can be computed in time polynomial in $n$. Note that the polynomial $F$ is explicit.

We now state our main correlation result using which we will obtain the lower bound against the Hadamard product of monotone multilinear polynomials in $\mathbb{C}[x_1, \ldots, x_n]$. A proof sketch is given in the appendix.

**Lemma 2.** *Let $F \in \mathbb{C}[x_1, \ldots, x_n]$ be the explicit multilinear polynomial defined above and $f_1, f_2 \in \mathbb{C}[x_1, \ldots, x_n]$ be any $1/3$-product polynomials. Then*

1. $\sum_{m \in \mathcal{M}(\{x_1, \ldots, x_n\})} F(m) \geq 0$.
2. $\mathrm{Corr}\,(F, f_1 \circ f_2) \leq 2^{-\alpha n} \|F\| \|f_1 \circ f_2\|$, *for a constant $\alpha > 0$ that is independent of $f_1$ and $f_2$.*

Using the above lemma bounding the correlation between $F$ and the Hadamard product of $1/3$-product polynomials, we will prove the main lower bound. We first recall a crucial lemma of Raz and Yehudayoff [RY08].

**Lemma 3.** *For $n \geq 3$, let $F \in \mathbb{C}[x_1, \ldots, x_n]$ be a monotone multilinear polynomial computed by a monotone circuit of size $s$ (i.e. the circuit has at most $s$ edges). Then, there are $s + 1$ monotone $1/3$-product polynomials $f_1, f_2, \ldots, f_{s+1}$ such that $F = \sum_{i=1}^{s+1} f_i$.*

**Theorem 9.** *For large enough $n \in \mathbb{N}$, there is an explicit monotone multilinear polynomial $F' \in \mathbb{Q}[x_1, \ldots, x_n]$ that cannot be written as a Hadamard product of two monotone multilinear polynomials $f_1, f_2 \in \mathbb{R}[x_1, \ldots, x_n]$ such that each $f_i$ is computed by monotone circuits of size less than $2^{\alpha n}$, where $\alpha > 0$ is an absolute constant.*

*Proof.* By the density of primes it suffices to consider $n$ of the form $tp$, for prime $p$, where $t$ is the constant in the definition of $F$. Let $X$ denote the set of variables $\{x_1, \ldots, x_n\}$, and let $F$ be the explicit polynomial mentioned in Lemma 2 above. For any monomial $m \in \mathcal{M}(X)$, let $F'(m) = (F(m) + 1)/2$. Clearly, the coefficients of $F'$ all lie in $\{0, 1\}$. Consider the correlation between $F$ and $F'$:

$$\mathrm{Corr}\,(F, F') = \left| \sum_{m : F(m) = 1} 1 \right| \geq 2^{n-1}$$

where the inequality above follows from the point 1 of Lemma 2.

Let us assume that $F'$ can be written as $f_1 \circ f_2$, where $f_1$ and $f_2$ are multilinear polynomials computed by monotone arithmetic circuits of size at most $s$. We assume $n \geq 3$, so that Lemma 3 is applicable. By Lemma 3, there exist monotone $1/3$-product polynomials $f_{1,1}, \ldots, f_{1,s+1}, f_{2,1}, \ldots, f_{2,s+1}$ such that $f_i = \sum_{j=1}^{s+1} f_{i,j}$, for each $i \in \{1, 2\}$. Thus, we have,

$$F' = \left( \sum_{j=1}^{s+1} f_{1,j} \right) \circ \left( \sum_{k=1}^{s+1} f_{2,k} \right) = \sum_{1 \leq j, k \leq s+1} f_{1,j} \circ f_{2,k}$$

Taking correlation with $F$ on both sides, we see that,

$$2^{n-1} \leq \sum_{1 \leq j, k \leq s+1} \mathrm{Corr}\,(F, f_{1,j} \circ f_{2,k}) \leq \sum_{1 \leq j, k \leq s+1} 2^{-\beta n} \|F\| \|f_{1,j} \circ f_{2,k}\|,$$

by applying triangle inequality and then part 2 of Lemma 2, where $\beta > 0$ is some constant.

Since, $f_{1,j} \circ f_{2,k}$'s are monotone polynomials adding up to $F'$, it follows that for any monomial $m \in \mathcal{M}(X)$ its coefficient in $f_{1,j} \circ f_{2,k}$ is at most 1. Hence, $\|f_{1,j} \circ f_{2,k}\| \leq \|F\|$ and we have

$$2^{n-1} \leq \sum_{1 \leq j,k \leq s+1} 2^{-\beta n} \|F\|^2 = (s+1)^2 2^{n-\beta n}$$

Consequently, we have $s \geq 2^{\beta n/4}$, for large enough $n$. ∎

## References

ABO99. E. ALLENDER, R. BEALS, AND M. OGIHARA, The complexity of matrix rank and feasible systems of linear equations, *Computational Complexity* , 8(2):99-126, 1999.

AO96. E. ALLENDER, M. OGIHARA, Relationships among PL, #L and the determinant. *RAIRO - Theoretical Informatics and Applications*, 30:1–21, 1996.

ARZ99. E. ALLENDER, K. REINHARDT, S. ZHOU, Isolation, matching and counting uniform and nonuniform upper bounds. *Journal of Computer and System Sciences*, 59(2):164–181, 1999.

AMS08. V. ARVIND, P. MUKHOPADHYAY, S. SRINIVASAN New results on Noncommutative Polynomial Identity Testing*In Proc. of Annual IEEE Conference on Computational Complexity,*268-279,2008.

Bh97. R. BHATIA, Matrix Analysis, Springer-Verlag Publishing Company, 1997.

Bo07. J. BOURGAIN: "On the Construction of Affine Extractors", Geometric & Functional Analysis GAFA, Vol. 17, No. 1. (April 2007), pp. 33-57.

BGK06. J. BOURGAIN, A. GLIBICHUK, S. KONYAGIN: "Estimate for the number of sums and products and for exponential sums in fields of prime order", J. London Math. Soc. 73 (2006), pp. 380-398.

GJ79. M. R. GAREY, D. S. JOHNSON Computers and Intractability: A Guide to the Theory of NP-Completeness. W.H. Freeman.*p. 228. ISBN 0-7167-1045-5*, 1979.

HMU. J. E. HOPCROFT, R. MOTAWANI, J. D. ULLMAN, Introduction to Automata Theory Languages and Computation,*Second Edition*, Pearson Education Publishing Company.

KI03. V. KABANETS, R. IMPAGLIAZZO, Derandomization of polynomial identity test means proving circuit lower bounds, *In Proc. of 35th ACM Sym. on Theory of Computing*,355-364,2003.

KvM02. A. KLIVANS, D. VAN MELKEBEEK, Graph nonisomorphism has subexponential size proofs unless the polynomial-time hierarchy collapses. *SIAM Journal on Computing*, 31(5):1501–1526, 2002.

N91. N. NISAN, Lower bounds for noncommutative computation *In Proc. of 23rd ACM Sym. on Theory of Computing,* 410-418, 1991.

RS05. R. RAZ, A. SHPILKA, Deterministic polynomial identity testing in non commutative models *Computational Complexity,*14(1):1-19, 2005.

RY08. RAN RAZ, AMIR YEHUDAYOFF, "Multilinear Formulas, Maximal-Partition Discrepancy and Mixed-Sources Extractors." FOCS 2008: 273-282.

T91. S. TODA, Counting Problems Computationally Equivalant to the Determinant, manuscript.

# Appendix

**Proof of Proposition 3**

First we prove the forward direction by constructing an acyclic CFG $G = (V, T, P, S)$ for mon($f$). Let $V = \{A_g|\ g\text{ is a gate of circuit } C\}$ be the set of nonterminals of $G$. We include a production in $P$ for each gate of the circuit $C$. If $g$ is an input gate with input $x_i, 1 \leq i \leq n$ include the production $A_g \to x_i$ in $P$. If the input is a *nonzero* field element then add the production $A_g \to \epsilon$.[3] $f_g$ be the polynomial computed at gate $g$ of $C$. If $g$ is a $\times$ gate with $f_g = f_h \times f_k$ then include the production $A_g \to A_h A_k$ and if it is $+$ gate with $f_g = f_h + f_k$ include the productions $A_g \to A_h \mid A_k$. Let the start symbol $S = A_g$, where $g$ is the output gate of $C$. It is easy to see from the above construction that $G$ is acyclic moreover $size(G) = O(s)$ and it generates the finite language mon($f$). The converse direction is similar.

**Proof of Theorem 8**

We first show that the complement of the problem is in NP. The NP machine will guess a monomial $m_\alpha \in X^*$, $X = \{x_1, \ldots, x_n\}$ and check if coefficient of $m_\alpha$ is nonzero in both $C$ and $C'$. Note that we can compute coefficient of $m_\alpha$ in $C$ and $C'$ in deterministic polynomial time using result from [AMS08]. Denote by CFGINT the problem of testing emptiness of the intersection of two acyclic CFGs that generate poly($n$) length strings. By Lemma 3 CFGINT is polynomial time many-one reducible to testing if $f \circ g$ is identically zero. The problem of testing if the intersection of two CFGs (with recursion) is empty is known to be undecidable via a reduction from the Post Correspondence problem [HMU, Chapter 9, Page 422]. We can give an analogous reduction from *bounded* Post Correspondence to CFGINT. The coNP-hardness of CFGINT follows from the coNP-hardness of bounded Post Correspondence [GJ79].

**Proof of Lemma 2**

Part 1 of Lemma 2 is trivial. By construction, each of the coefficients of $F$ is $\pm 1$. For any $z \in \mathbb{F}$ note that $\sum_{y \in \mathbb{F}} \psi(z \cdot y) \geq 0$. Hence,

$$\sum_{m \in \mathcal{M}(X)} F(m) = \sum_{y_1, \ldots, y_t \in \mathbb{F}} \psi(\prod_{j=1}^{t} y_j) = \sum_{y_1, \ldots, y_{t-1}} \sum_{y_t} \psi((\prod_{j=1}^{t-1} y_j) y_t) \geq 0.$$

where the last inequality follows as each of the terms in the outer summation is non-negative.

---

[3] If the circuit takes as input 0, we can first propagate it through the circuit and eliminate it.

**The exponential sum estimate** We now state an exponential sum estimate of Bourgain, Glibichuk, and Konyagin (see [BGK06],[Bo07]) that we will need later. The result is a special case of their result, and is similar to the version used in [RY08].

**Theorem 10.** *There exist two constants, an integer $s \in \mathbb{N}$ and $\gamma > 0$, such that for every prime $p$, for every family of sets $A_1, A_2, \ldots, A_s \subseteq \mathbb{F}_{2^p}$ of size at least $2^{p/20}$ each, for every nonzero $z \in \mathbb{F}_{2^p}$, and for each non-trivial additive character $\psi$ of $\mathbb{F}_{2^p}$,*

$$\left| \sum_{y_1 \in A_1, \ldots, y_s \in A_s} \psi\left(z . \prod_{i=1}^{s} y_i\right) \right| \leq 2^{-\gamma p} |A_1|.|A_2| \ldots |A_s|$$

We use the above fixed $s \in \mathbb{N}$ as the constant $s$ in the construction of the polynomial $F$ of Lemma 2.

**A strengthening of the result of [RY08]** We can prove, exactly along the lines of [RY08, Theorem 3.1], that the polynomial $F$ has low correlation with $1/3$-product polynomials. However, we need to prove the stronger claim that it has low correlation with Hadamard products of such polynomials. In order to prove this claim we need to strengthen [RY08, Theorem 3.1].

Let $X = X' \cup X''$ be a partition of the variable set $X$. For a monomial $m'' \in \mathcal{M}(X'')$, we call the tuple $(X', X'', m'')$ a *suitable restriction* if for each $i \in [t]$ such that $|X'' \cap X(i)| \geq p/2$, some variable in $X(i)$ appears in $m''$. By our encoding assumption, this implies that for any monomial $m' \in \mathcal{M}(X')$ and any $i$ such that $|X'' \cap X(i)| \geq p/2$, $y_i(m'm'') \neq 0$.

Given a suitable $(X', X'', m'')$, denote by $\tilde{F}$ the multilinear polynomial over the variables $X'$ where, for any monomial $m' \in \mathcal{M}(X')$, $\tilde{F}(m') = F(m'm'')$. Let $f = gh \in \mathbb{C}[X']$ be a multilinear polynomial with $g \in \mathbb{C}[A]$ and $h \in \mathbb{C}[B]$, where $A$ and $B$ are disjoint sets. The required stronger version of [RY08, Theorem 3.1] is the following.

**Theorem 11.** *Assume $(X', X'', m'')$ is suitable restriction. Let $\tilde{F}$ be defined as above and let $f = gh$ as above with $|A|, |B| \geq n/10$. Then,*

$$\mathrm{Corr}\left(\tilde{F}, f\right) \leq \frac{\|\tilde{F}\| \|f\|}{2^{\Omega(n)}}$$

*where the constant in the $\Omega(\cdot)$ is independent of $f$.*

*Proof Sketch.* Our notation is from [RY08]. For $i \in [t]$, let $A(i)$ and $B(i)$ denote $A \cap X(i)$ and $B \cap X(i)$ respectively. We need the following simple claim, the proof of which is similar to [RY08, Proposition 9.2].

*Claim.* There are at least $s+1$ many $i \in [t]$ such that $|A(i)| \geq p/20$ and at least $s+1$ many $j \in [t]$ such that $|B(j)| \geq p/20$.

Fix $I \subseteq [t]$ of size $s$ such that $|A(i)| \geq p/20$ for each $i \in I$. Let $J = [t] \setminus I$. By the above claim, there is a $j_0 \in J$ such that $|B(j_0)| \geq p/20$.

Set $A_1 = \bigcup_{i \in I} A(i), B_1 = \bigcup_{j \in J} B(j)$ and $A_2 = A \setminus A_1, B_2 = B \setminus B_1$. We denote by $a_1, a_1'$ etc. monomials from $\mathcal{M}(A_1)$ and similarly for monomials from $\mathcal{M}(A_2), \mathcal{M}(B_1)$ and $\mathcal{M}(B_2)$. Finally, we denote by $m_1$ and $m_2$ the restriction of the monomial $m''$ to the sets $\bigcup_{i \in I} X(i)$ and $\bigcup_{j \in J} X(j)$ respectively.

Given monomials $a_2 \in \mathcal{M}(A_2)$ and $b_1, b_1' \in \mathcal{M}(B_1)$, denote by $Z(a_2, b_1, b_1')$ the field element $\prod_{j \in J} y_j(a_2 b_1 m_2) - \prod_{j \in J} y_j(a_2 b_1' m_2)$. Let $S(a_2)$ denote those pairs $(b_1, b_1')$ such that $Z(a_2, b_1, b_1') = 0$. Let $S_1 = \left\{ a_2 \mid |S(a_2)| > 2^{2|B_1| - p/40} \right\}$ and $S_2 = \left\{ a_2 \mid |S(a_2)| \leq 2^{2|B_1| - p/40} \right\}$.

The quantity we wish to bound is:

$$\mathrm{Corr}\left(\tilde{F}, f\right) = \left| \sum_{a_1, a_2, b_1, b_2} \tilde{F}(a_1 a_2 b_1 b_2) \overline{f(a_1 a_2 b_1 b_2)} \right|$$

$$\leq \underbrace{\left| \sum_{\substack{a_2 \in S_1 \\ a_1, b_1, b_2}} \tilde{F}(a_1 a_2 b_1 b_2) \overline{f(a_1 a_2 b_1 b_2)} \right|}_{C_1} + \underbrace{\left| \sum_{\substack{a_2 \in S_2 \\ a_1, b_1, b_2}} \tilde{F}(a_1 a_2 b_1 b_2) \overline{f(a_1 a_2 b_1 b_2)} \right|}_{C_2}$$

(3)

We first bound $C_1$ using the following analogue of [RY08, Corollary 9.6].

*Claim.* For large enough $p$, $|S_1| \leq 2^{|A_2| - p/50}$.

*Proof of Claim.* Let $S = \{(a_2, b_1, b_1') \mid Z(a_2, b_1, b_1') = 0\}$. We will bound $|S|$. For this we first bound the number of $(a_2, b_1, b_1')$ such that $\prod_{j \in J} y_j(a_2 b_1' m_2) = 0$.

Fix a $j \in J$. If $j$ is such that $|X(j) \cap X''| \geq p/2$ then, as $(X', X'', m'')$ is a suitable restriction, we have $y_j(a_2 b_1' m_2) \neq 0$. Otherwise $|X(j) \cap X'| \geq p/2$, and $y_j(a_2 b_1' m_2) = 0$ only if none of the variables in $X(j) \cap X'$ appears in $a_2$ or $b_1'$; the number of such triples $(a_2, b_1, b_2')$ is at most $2^{|A_2| + 2|B_1| - p/2}$. Thus, the number of $(a_2, b_1, b_1')$ such that $\prod_{j \in J} y_j(a_2 b_1' m_2) = 0$ is at most $t 2^{|A_1| + 2|B_2| - p/2}$.

If $\prod_{j \in J} y_j(a_2 b_1' m_2) \neq 0$ then $Z(a_2, b_1, b_2') = 0$ only if

$$y_{j_0}(a_2 b_1' m_2) = \frac{\prod_{j \in J} y_j(a_2 b_1 m_2)}{\prod_{j \in J \setminus \{j_0\}} y_j(a_2 b_1' m_2)}.$$

I.e. the scalar $y_{j_0}(a_2 b_1' m_2)$, and hence the restriction of $b_1'$ to $X_{j_0}$, is completely determined by $a_2, b_1$, and the restriction of $b_1'$ to $X \setminus X_{j_0}$. Since $|B(j_0)| \geq p/20$, the number of $(a_2, b_1, b_1')$ such that this is true is at most $2^{|A_2|+2|B_1|-p/20}$. Hence,

$$|S| \leq t 2^{|A_2|+2|B_1|-p/2} + 2^{|A_2|+2|B_1|-p/20} \leq 2^{|A_2|+2|B_1|-p/20+1}$$

for large enough $p$. On the other hand, $|S|$ is at least $|S_1| \cdot 2^{2|B_1|-p/40}$. Hence, for large enough $p$ we have

$$|S_1| \leq 2^{|A_1|-p/40+1} < 2^{|A_1|-p/50}.$$

It follows from the above claim, and a simple application of the Cauchy-Schwarz inequality, that $C_1 \leq 2^{-\Omega(n)} \|\tilde{F}\| \|f\|$. Now we bound $C_2$. In this case, the proof of [RY08, Proposition 9.4] goes through almost verbatim to yield the bound; the only change necessary is in the proof of [RY08, Claim 9.7], where we need to use the more general exponential sum estimate stated in Theorem 10 above. We omit this proof and simply state the obtained bound on $C_2$.

$$C_2 \leq 2^{-\Omega(n)} \|\tilde{F}\| \|f\|.$$

Putting this together with Equation (3) and the bound on $C_1$ yields the statement of the theorem. ■

**The correlation bound** We now consider part 2 of Lemma 2. Let $f = f_1 \circ f_2$ be a Hadamard product of 1/3-product polynomials $f_1$ and $f_2$. For $i \in \{1, 2\}$, let $f_i = g_i h_i$, where $g_i \in \mathbb{C}[A_i], h_i \in \mathbb{C}[B_i]$; $A_i$ and $B_i$ being disjoint subsets of $X$ of size at least $n/3$ each. We can assume that $A_i \cup B_i = X$. Consider the four pairwise intersections $A_1 \cap A_2, A_1 \cap B_2, A_2 \cap B_1$, and $B_1 \cap B_2$. We can write any monomial $m \in \mathcal{M}(X)$ as $m_{11} m_{12} m_{21} m_{22}$, where $m_{11}, m_{12}, m_{21}$, and $m_{22}$ denote the restriction of $m$ to $A_1 \cap A_2, A_1 \cap B_2, A_2 \cap B_1$, and $B_1 \cap B_2$ respectively.

If $A_1 = A_2$ and $B_1 = B_2$ or $A_1 = B_2$ and $A_2 = B_1$, then the polynomial $f$ is a 1/3-product polynomial. Following [RY08] we can show that $F$ has very low correlation with all 1/3-product polynomials. Hence, in this easy case we are done. In the next claim we argue that an "approximate" version of this desirable scenario always holds.

*Claim.* Let $A_i, B_i$, for $i \in \{1, 2\}$ be as defined above. At least one of the following holds.
Case 1: $|A_1 \cap A_2| \geq n/10$ and $|B_1 \cap B_2| \geq n/10$.
Case 2: $|A_1 \cap B_2| \geq n/10$ and $|A_2 \cap B_1| \geq n/10$.

*Proof of Claim.* Assume Case 1 does not hold. Let us assume that $|A_1 \cap A_2| < n/10$ (the other case is symmetric to this one). Then, since $|A_2| \geq n/3$, we know that $|A_2 \cap B_1| = |A_2| - |A_2 \cap A_1|$ (this is because we have assumed that $A_1 \cup B_1 = X$), which is at least $n/3 - n/10 > n/10$. Similarly, $|A_1 \cap B_2|$ is also at least $n/10$. Thus, Case 2 holds.

By swapping the names of $A_2$ and $B_2$ if necessary, we assume that Case 1 of claim 4 holds. Let $X' = (A_1 \cap A_2) \cup (B_1 \cap B_2)$ and $X'' = X \setminus X' = (A_1 \cap B_2) \cup (A_2 \cap B_1)$.

We now note that, restricted to the set of variables $X'$, the polynomial $f$ has a 'product polynomial structure'. More precisely, for a monomial $m = m_{11}m_{12}m_{21}m_{22} \in \mathcal{M}(X)$, we can write $f(m)$ as the product of $g_1(m_{11}m_{12})g_2(m_{11}m_{21})$ and $h_1(m_{21}m_{22})h_2(m_{12}m_{22})$; for a fixed $m_{12}, m_{21}$, the former depends only on the monomial $m_{11}$ and the latter only on $m_{22}$: this is very much like a product polynomial. We use this further below. From now, we denote by $g_{12}(m_{11}m_{12}m_{21})$ and $h_{12}(m_{12}m_{21}m_{22})$ the values $g_1(m_{11}m_{12})g_2(m_{11}m_{21})$ and $h_1(m_{21}m_{22})h_2(m_{12}m_{22})$. Hence, we have,

$$\text{Corr}(F, f) = \left| \sum_{m_{11}, m_{12}, m_{21}, m_{22}} F(m_{11}m_{12}m_{21}m_{22}) \overline{f(m_{11}m_{12}m_{21}m_{22})} \right|$$

$$= \left| \sum_{m_{12}, m_{21}} \sum_{m_{11}, m_{22}} F(m_{11}m_{12}m_{21}m_{22}) \overline{g_{12}(m_{11}m_{12}m_{21}) h_{12}(m_{12}m_{21}m_{22})} \right| \tag{4}$$

For fixed $m_{12}$ and $m_{21}$, the inner summation above is the inner product of vectors that correspond to polynomials over the variables in $X'$: one of the vectors is the restriction of $F$ to these coordinates, and the other is the product $g_{12}h_{12}$, which is a $1/10$-product polynomial over the variables in $X'$. Our aim is to show that, for 'most' values of $m_{12}$ and $m_{21}$, the inner summation is small (to prove this, we will use the proof of [RY08]); for other values of $m_{12}$ and $m_{21}$, a brute force bound will do. We show this first.

Call a tuple of monomials $(m_{12}, m_{21}) \in \mathcal{M}(A_1 \cap B_2) \times \mathcal{M}(A_2 \cap B_1)$ a *suitable pair* of monomials if the tuple is $(X', X'', m_{12}m_{21})$ is a suitable restriction as defined in the previous section. Let $\mathcal{B}$ denote the set $\{(m_{12}, m_{21}) \in \mathcal{M}(A_1 \cap B_2) \times \mathcal{M}(A_2 \cap B_1) \mid (m_{12}, m_{21}) \text{ not suitable}\}$ of unsuitable pairs of monomials, and let $\mathcal{B}'$ denote the set of suitable pairs of monomials. We split the summation in Equation (4) as follows:

$$\text{Corr}(F, f) \leq \underbrace{\left| \sum_{(m_{12}, m_{21}) \in \mathcal{B}} \sum_{m_{11}, m_{22}} F(m_{11}m_{12}m_{21}m_{22}) \overline{g_{12}(m_{11}m_{12}m_{21})h_{12}(m_{12}m_{21}m_{22})} \right|}_{T_1} +$$

$$\underbrace{\left| \sum_{(m_{12}, m_{21}) \in \mathcal{B}'} \sum_{m_{11}, m_{22}} F(m_{11}m_{12}m_{21}m_{22}) \overline{g_{12}(m_{11}m_{12}m_{21})h_{12}(m_{12}m_{21}m_{22})} \right|}_{T_2} \quad (5)$$

We bound the sums $T_1$ and $T_2$ separately. We tackle $T_1$ first. We need a claim bounding the number of unsuitable pairs:

*Claim.* $|\mathcal{B}| \leq \frac{2^{|X''|}}{2^{\Omega(n)}}$, for large enough $n$.

*Proof of Claim.* Let $S \subseteq [t]$ be the set of those $i$ s.t $|X(i) \cap X''| \geq p/2$. For $i \in S$, let $\mathcal{B}_i$ denote those pairs of monomials $(m_{12}, m_{21}) \in \mathcal{M}(A_1 \cap B_2) \times \mathcal{M}(A_2 \cap B_1)$ such that no variable $x \in X(i) \cap X''$ appears in them.

Clearly, for $i \in S$, $|\mathcal{B}_i| \leq 2^{|X'' \setminus X(i)|} \leq 2^{|X''|-p/2}$. Also, since $\mathcal{B} = \bigcup_{i \in S} \mathcal{B}_i$, we have $|\mathcal{B}| \leq |S| 2^{|X''|-p/2} \leq t 2^{|X''|-p/2}$. Since $p = n/t = \Omega(n)$, we see that $|\mathcal{B}| \leq 2^{|X''|}/2^{\Omega(n)}$, for large enough $n$.

We now bound $T_1$. Using the Cauchy-Schwarz inequality, we have,

$$T_1 \leq \sqrt{\sum_{(m_{12}, m_{21}) \in \mathcal{B}} \sum_{m_{11}, m_{22}} |F(m_{11}m_{12}m_{21}m_{22})|^2} \cdot$$

$$\sqrt{\sum_{(m_{12}, m_{21}) \in \mathcal{B}} \sum_{m_{11}, m_{22}} |g_{12}(m_{11}m_{12}m_{21})h_{12}(m_{12}m_{21}m_{22})|^2}$$

$$\leq \sqrt{|\mathcal{B}| 2^{|X''|}} \cdot \sqrt{\sum_{m_{12}, m_{21}, m_{11}, m_{22}} |g_{12}(m_{11}m_{12}m_{21})h_{12}(m_{12}m_{21}m_{22})|^2}$$

$$\leq 2^{\frac{|X'|+|X''|}{2} - \Omega(n)} \cdot \sqrt{\sum_m |f(m)|^2} \leq \frac{\|F\|}{2^{\Omega(n)}} \|f\| = 2^{-\Omega(n)} \|F\| \|f\| \quad (6)$$

Above, we have used the fact that $|F(m)| = 1$ for all monomials $m$, and that for $m = m_{11}m_{12}m_{21}m_{22}$, $f(m) = g_{12}(m_{11}m_{12}m_{21})h_{12}(m_{12}m_{21}m_{22})$.

We now bound $T_2$. Fix any suitable pair of monomials $(m_{12}, m_{21})$. For $m_{11} \in \mathcal{M}(A_1 \cap A_2)$ and $m_{22} \in \mathcal{M}(B_1 \cap B_2)$, we denote by $\tilde{F}(m_{11}m_{22})$, $\tilde{g}(m_{11})$ and $\tilde{h}(m_{22})$ the values $F(m_{11}m_{12}m_{21}m_{22})$, $g_{12}(m_{11}m_{12}m_{21})$, and $h_{12}(m_{12}m_{21}m_{22})$ respectively. We think of $\tilde{F}, \tilde{g}$, and $\tilde{h}$ as vectors of multilinear polynomials over $X'$. Looked at in this way, $\tilde{g} \in \mathbb{C}[A_1 \cap A_2]$ and $\tilde{h} \in \mathbb{C}[B_1 \cap B_2]$. By Claim 4, $|A_1 \cap A_2|, |B_1 \cap B_2| \geq n/10$. Hence, Theorem 11 is applicable, and we have

$$\operatorname{Corr}\left(\tilde{F}, \tilde{g}\tilde{h}\right)^2 \leq 2^{|X'|-\Omega(n)}\|\tilde{g}\tilde{h}\|^2 \tag{7}$$

Using the above bound, we bound $T_2$ and finish the proof of Lemma 2.

Squaring the expression for $T_2$ in Equation 5, and using the Cauchy-Schwarz inequality, we have,

$$\begin{aligned}
T_2^2 &\leq |\mathcal{B}'| \sum_{(m_{12},m_{21})\in\mathcal{B}'} \left|\sum_{m_{11},m_{22}} F(m_{11}m_{12}m_{21}m_{22})\overline{g_{12}(m_{11}m_{12}m_{21})h_{12}(m_{12}m_{21}m_{22})}\right|^2 \\
&\leq |\mathcal{B}'| 2^{|X'|-\Omega(n)} \sum_{(m_{12},m_{21})\in\mathcal{B}'} \sum_{m_{11},m_{22}} |g_{12}(m_{11}m_{12}m_{21})h_{12}(m_{12}m_{21}m_{22})|^2 \quad \text{(by Equation 7)} \\
&\leq 2^{|X''|+|X'|-\Omega(n)} \sum_{(m_{12},m_{21})\in\mathcal{B}'} \sum_{m_{11},m_{22}} |f(m_{11}m_{12}m_{21}m_{22})|^2 \quad (\because |\mathcal{B}'| \leq 2^{|X''|}) \\
&\leq 2^{n-\Omega(n)} \sum_{m\in\mathcal{M}(X)} |f(m)|^2 \leq \frac{\|F\|^2}{2^{\Omega(n)}}\|f\|^2. \quad \text{(since } \|F\|^2 = 2^n\text{)}
\end{aligned}$$

$\therefore T_2 \leq 2^{-\Omega(n)}\|F\|\|f\|.$

Using the above bound on $T_2$, and Equations 4 and 6, we see that

$$\operatorname{Corr}(F, f) \leq T_1 + T_2 \leq 2^{-\Omega(n)}\|F\|\|f\|.$$

This proves part 2 of Lemma 2.